# An experimental demonstration of room-temperature spin transport in n-type Germanium epilayers

S. Dushenko[1†], M. Koike[1†], Y. Ando[1,2], T. Shinjo[2], M. Myronov[3], and M. Shiraishi [1,2*]

1. Graduate School of Engineering Science, Osaka University, Toyonaka 560-8531, Japan.

2. Department of Electronic Science and Engineering, Kyoto University, Kyoto 615-8510, Japan.

3. Department of Physics, The University of Warwick, Coventry CV4 7AL, United Kingdom.

†These authors contributed equally to this work.

* Corresponding author: Masashi Shiraishi (mshiraishi@kuee.kyoto-u.ac.jp)

We report the first experimental demonstration of room-temperature spin transport in n-type Ge epilayers grown on a Si(001) substrate. By utilizing spin pumping under ferromagnetic resonance, which inherently endows a spin battery function for **semiconductors connected with the ferromagnet, a pure spin current is generated in the n-Ge at room temperature. The pure spin current is detected by using the inverse spin Hall effect of either Pt or Pd electrode on the n-Ge. A theoretical model including a geometrical contribution allows to estimate a spin diffusion length in n-Ge at room temperature to be 660 nm. The temperature dependence of the spin relaxation time provides evidence for Elliott-Yafet spin relaxation mechanism.**

Group IV semiconductors, such as Si and Ge, continue to attract tremendous attention in spintronics due to their suppression of the spin-orbit interaction (SOI), i.e., the suppression of spin relaxation. Crystal inversion symmetry of Si and Ge precludes the spin relaxation of conduction electrons by the D'yakonov-Perel mechanism, resulting in long spin relaxation time. Although the SOI of Ge is not negligibly small (0.29 eV), the SOI affects electrons much weaker than holes in Ge, making n-type Ge a promising material for spin transport. Furthermore, Ge possesses much higher carrier mobility comparing to Si [1], thus it is possible for Ge-based spin transport field effect transistors with a small gate length to overcome scaling limits of the Si-based devices. In fact hole mobility at low temperatures in Ge recently was reported to be higher than 1000000 $cm^2V^{-1}s^{-1}$- [2]. Additionally, in recent years significant success was achieved in the production of high quality $GeO_2$ layer with gate function in the n-channel Ge based metal oxide semiconductor field effect transistor [3]. However, in spite of all recent progress in Ge field and reports on successful spin injection at room temperature (RT), in contrast to Si, spin transport in Ge was observed only at low temperatures up to date [4-9]. Whereas in highly doped n-type [10] and p-type [11] Ge, the inverse spin Hall effect [12-14] (ISHE) was recently observed at RT; it is notable that no direct observations of pure spin transport have been reported. Thus realization of the RT spin transport in Ge is long awaited for further progress in semiconductor spintronics.

The spin-pumping-induced generation of pure spin current originates from magnetization ($\mathbf{M}(t)$) precession in the ferromagnetic layer under ferromagnetic resonance (FMR) conditions [15-18]. Using this very potential tool successful spin transport, has been achieved at RT in p-type Si [19], single-layer graphene [20] and semiconducting conjugated polymers [21]. In our experiment, the magnetic moment is transferred through the interface of the ferromagnet into the adjacent n-Ge layer, creating pure spin current in the latter (Fig. 1(a)). After the propagation through the n-Ge channel, the spins are absorbed in the metal electrode with strong SOI $Me^{SOI}$ [22] (Pt and Pd in our study). The SOI in $Me^{SOI}$ converts the spin current into charge current via the ISHE [23-30], which is described by the following relation [31]:

$\mathbf{J_C} = D_{ISHE}\mathbf{J_S} \times \boldsymbol{\sigma}$, where $D_{ISHE}$ represents the ISHE efficiency of the material. The generated charge current can be detected as a voltage at the ends of the Me$^{SOI}$ strip.

While electronic properties of Ge are superior to those of Si, Ge wafers are heavier, less durable and much more expensive than its Si analogies. So it is desirable to combine high mobility of Ge channels with durability and low prices of Si wafers. However, lattice mismatch of 4.2% between Si and Ge ($a_0^{Si} = 5.431$ Å and $a_0^{Ge} = 5.657$ Å) precludes direct epitaxial growth of high quality relaxed Ge layers on top of the Si wafers. Efforts to overcome this difficulty using different approaches are still ongoing in Ge related research [32-35]. In our study high quality Ge channel was successfully grown on top of the Si (001) substrate using the two temperature method. Additionally, in contrast to the usual use of solid-source molecular beam epitaxy and low-energy plasma-enhanced chemical vapor deposition for production of the high quality channels, our channel was produced using the reduced pressure chemical vapor deposition technique (RP-CVD) highly suitable for industrial production. This makes our study a ready-to-use for applications.

Figure 1(a) shows the structure of the n-Ge-based spin transport device used in our study. The Ge epilayers were grown on standard p⁻-Si(001) substrates using two temperature growth method by RP-CVD [33,35]. The structure consisted of an 1-µm-thick undoped Ge epilayer and 50 nm heavily n-type doped Ge epilayer with a phosphorous doping concentration of ~1.0×10$^{19}$ cm$^{-3}$ and a degree of relaxation of 104%, calculated from analysis of measured high resolution X-ray diffraction symmetrical and asymmetrical reciprocal space maps. This over relaxation of the Ge channel is attributed to the difference in the thermal expansion coefficients between Ge and Si ($\Delta a/(a\Delta T)|_{Si,\Delta T=1°C} = 5.8 \times 10^{-6}(°C)^{-1}$ and $\Delta a/(a\Delta T)|_{Ge,\Delta T=1°C} = 2.6 \times 10^{-6}(°C)^{-1}$ [1]), i.e. Ge channel is 100% relaxed during the temperature growth, but under slight tensile strain after cooling down to the RT [33-34,36-37]. The Ge epilayers were measured to have root mean square roughness below 1 nm by atomic force microscopy (AFM) and threading dislocation density of $\sim 5 \times 10^6$ cm$^{-2}$, these very low values confirms exceptional quality of the produced n-Ge epilayers. Me$^{SOI}$ and Ni$_{80}$Fe$_{20}$ (Py) strips were formed on top of the n-Ge epilayers by electron beam lithography and electron beam evaporation. The samples were etched in 10% hydrofluoric acid solution and washed in deionized water prior to the evaporation of Me$^{SOI}$ to remove the natural Ge oxide layer.

Hereafter, we refer to this sample type as Py/n-Ge/Me[SOI]. Samples were placed in the center of a TE$_{102}$ cavity inside an electron spin resonance system with a microwave frequency $f = 9.58$ GHz. Ag paste was used to attach one Cu wire at each end of the Me[SOI] strip to detect voltage signal. Using identical procedures, three different sample types were prepared, namely, with a Cu strip instead of the Me[SOI] strip (Py/n-Ge/Cu in Fig. 1(b)), with Cu and Pd strips on different sides of the Py (Cu/n-Ge/Py/n-Ge/Pd in Fig. 1(c)) and with only Py on top of the thermal silicon oxide (Py/SiO$_2$ in Fig. 1(d)). The first two sample types were used for the control experiments, while the third was used to calculate the spin current density at the Py/n-Ge interface, as explained in detail in Supplemental Material. Measurements, when not mentioned explicitly, were carried out at RT.

Now we proceed to experimental results. Figure 2(a) shows the first derivative of the FMR spectrum, d$I$/d$H$. The red and black lines represent the spectra for Py/SiO$_2$ and Py/n-Ge/Pd, respectively. The enhanced peak-to-peak width of the FMR signal in the second case is due to the induction of spin pumping from Py into n-Ge under the FMR conditions. The ISHE voltage is proportional to the generated spin current, the amplitude of which is proportional to the microwave absorption, which is maximized at the resonance field $H_{FMR}$. Hence, the voltage signal from the ISHE takes the shape of a symmetric peak with respect to $H_{FMR}$. The detected electromotive force was fitted using the function [31] $V(H) = V_{ISHE} \frac{\Gamma^2}{(H-H_{FMR})^2+\Gamma^2} + V_{asym} \frac{-2\Gamma(H-H_{FMR})}{(H-H_{FMR})^2+\Gamma^2} + aH + b,$ where the first term describes the symmetric contribution to the voltage signal from the ISHE. The second term describes the asymmetric contribution to the voltage from different spurious effects, including the AHE, which voltage sign is reversed at the $H_{FMR}$; additionally, the last two terms represent the offset voltage (Fig. 2(b)). Fitting the experimentally detected voltage yields the values of $V_{ISHE}$ = 1.73 µV and $V_{asym}$ = -0.43 µV. Finally, to eliminate any heating effects average of the $V_{ISHE}$ for opposing orientations of the external magnetic field, **H**, $\theta_H = 0°$ (shown on Fig. 1(a)) and $\theta_H = 180°$, was calculated as $V_{ISHE} = (V_{ISHE}|_{\theta_H=0°} - V_{ISHE}|_{\theta_H=180°})/2,$ giving a value of $V_{ISHE}$ = 1.62 µV. Figures 3(a)-3(b) show an analogous symmetrical in shape voltage signal detected from the Py/n-Ge/Pt sample.

The $V_{ISHE}$ is proportional to the square of the microwave magnetic field, *h*, making $V_{ISHE}$ linearly proportional to the microwave power, $P_{MW}$, as $V_{ISHE} \propto h^2 \propto P_{MW}$ [18, 25]. Consistent with this prediction is the fact that $V_{ISHE}$ increased linearly with the $P_{MW}$ for both the Py/n-Ge/Pt and Py/n-Ge/Pd samples (Figs. 3(c)-3(d)). The reversal of the external magnetic field, **H**, to the opposite direction causes σ to change sign, which in turn leads to a change in the sign of the ISHE electric field, $\mathbf{E}_{ISHE} \propto \mathbf{J}_C$. Figures 3(a)-3(b) demonstrate the change in sign of the electromotive force upon the reversal of the magnetic field, **H**, to the opposite direction, thus demonstrating that the relation $\mathbf{J_C} = D_{ISHE}\mathbf{J_S} \times \boldsymbol{\sigma}$ for the ISHE holds in our system.

Next, we theoretically calculate the amplitude of the $V_{ISHE}$ for the Py/n-Ge/Pd sample by expanding the simplest model used previously with taking a geometric effect into account [19]. The theoretical model used in this study allows more precise estimation of spin coherence length. Estimation of the real part of the mixing conductance $g_r^{\uparrow\downarrow}$ and spin current density at the ferromagnet/nonferromagnet interface is well established in the number of papers [25-27]. We calculate them to be $g_r^{\uparrow\downarrow} = 2.15 \times 10^{19}$ m$^{-2}$ and the spin current density at the Py/n-Ge interface $j_S^{Py/n-Ge} = 1.33 \times 10^{-9}$ Jm$^{-2}$ (see Supplemental Material for details). Further we propose simple geometrical model to take into account spin current dissipation in the n-Ge channel. During spin transport from Py to Pd through the n-Ge channel, the density of the spin current $j_S^{Py/n-Ge}$ is exponentially damped on the spin diffusion length $\lambda_{n-Ge}$ of the n-Ge. Taking into account our device geometry, we assumed that half of the Py strip contributes to the spin current in the direction of the Pd strip. Integrating over this half gives the spin current density at the n-Ge/Pd interface as

$$j_S^{n-Ge/Pd} = j_S^{Py/n-Ge} \frac{1}{w_{Pd}} \int_0^{\frac{w_{Py}}{2}} e^{-\frac{L_{Py-Pd}+x}{\lambda_{n-Ge}}} dx = j_S^{Py/n-Ge} e^{\frac{-L_{Py-Pd}}{\lambda_{n-Ge}}} \frac{\lambda_{n-Ge}}{w_{Pd}} \left(1 - e^{-\frac{w_{Py}}{2\lambda_{n-Ge}}}\right), \quad (1)$$

where $w_{Pd} = 1.5$ μm is the width of the Pd strip. The gap length $L_{Py-Pd}$ was measured to be 620 nm using AFM. The conductivity of the Pd $\sigma_{Pd} = 1.97 \times 10^6$ (Ωm)$^{-1}$ [25] is more than one order of magnitude higher than that of our n-Ge channel $\sigma_{n-Ge} = 8.22 \times 10^4$ (Ωm)$^{-1}$; this fact and calculated $j_S^{n-Ge/Pd}$ allows us to modify commonly used expression [19,27] and finally write down the voltage of the ISHE from the Pd strip as

$$V_{ISHE} = \frac{l_{Py} \theta_{SHE} \lambda_{Pd} \tanh(d_{Pd}/2\lambda_{Pd})}{d_{Pd} \sigma_{Pd}} \left(\frac{2e}{\hbar}\right) j_S^{n-Ge/Pd}, \quad (2)$$

where $l_{Py} = 900$ μm is the length of the Py strip, $d_{Pd} = 10$ nm is the thickness of the Pd strip, and $\theta_{SHE} = 0.01$ [25] and $\lambda_{Pd} = 9$ nm [23] are the spin-Hall angle and the spin diffusion length of the Pd, respectively. By equating (2) to the experimentally measured $V_{ISHE} = 1.62$ μV, we calculated the value of the spin diffusion length in the n-Ge to be $\lambda_{n-Ge} = 680$ nm.

Formation of germanide at the n-Ge/Me[SOI] interface can take place at comparatively low temperatures, which can be achieved during the metal evaporation process; this may have led to the changes in spin transport properties of the n-Ge/Me[SOI] interface. However, these changes do not depend on the gap distance $L_{Py-Me^{SOI}}$ between the Py and Me[SOI] strips. To rectify the $\lambda_{n-Ge}$ estimation, the dependence of the gap length $L_{Py-Me^{SOI}}$ was measured. The $V_{ISHE}$ behavior over changes in $L_{Py-Me^{SOI}}$ is governed by Eq. (1) through spin current density damping in the n-Ge layer. Figure 4(a) shows the fitting of the gap length $L_{Py\text{-}Pt}$ dependence of the normalized $V_{ISHE}$ with function

$$V_{ISHE}/j_S^{Py/n-Ge} = A\lambda_{n-Ge}e^{-L/\lambda_{n-Ge}}\left(1 - e^{-\frac{w_{Py}}{2\lambda_{n-Ge}}}\right). \tag{3}$$

By using the fitting function (3), the refined estimation of the spin diffusion length $\lambda_{n-Ge}$ is calculated to be 460 nm. From the gap length $L_{Py\text{-}Pd}$ dependence of the normalized $V_{ISHE}$ in the Py/n-Ge/Pd samples (Fig. 4(b)), $\lambda_{n-Ge}$ is calculated to be 576 nm. Finally, taking the data from all samples into account, we estimate the spin diffusion length in the n-Ge channel at RT to be $\lambda_{n-Ge} = 660 \pm 200$ nm. The spin diffusion length in highly doped n-Ge (with a doping concentration of $10^{18}$ - $2\times10^{19}$ cm$^{-3}$) was reported to be 580 nm at 4 K [4], 590 nm at 150 K [9], 683 nm and 1300±200 nm at RT ([5] and [7], respectively). The discrepancy in the value of $\lambda_{n-Ge}$ is mainly derived from the difference in the quality and features of the n-Ge channel used, as well as from the chosen experimental method (three-terminal [5,7] and nonlocal [4,8,9] Hanle measurements). However, we stress that none of these experiments demonstrated RT spin transport in n-Ge, and, to the extent of our knowledge, we are the first to show clear evidence of successful spin transport at RT in n-Ge.

We measured temperature dependence of the ISHE voltage in the Py/n-Ge/Pt sample to shed light on the spin relaxation mechanism in our n-Ge epilayers. We assume the Elliott-Yafet mechanism of spin Hall conductivity of Pt, i.e. $\lambda_{Pd}(T) \propto \sigma_{Pd}(T)$ and the constant spin Hall conductivity with changing temperature [38-40]. Using aforementioned we extract $\lambda_{n-Ge}(T)/\lambda_{n-Ge}(297\ K)$ from Eq. (2) (Fig. 5(a)). Also temperature dependence of mobility of the n-Ge channel $\mu_{n-Ge}$ was determined using maximum-entropy mobility spectrum analysis [41]. Mobility in the n-Ge channel changed from 210±30 cm$^2$V$^{-1}$s$^{-1}$ at RT to 376±30 cm$^2$V$^{-1}$s$^{-1}$ at 130 K (see Supplemental Material Fig. S1 for $\mu_{n-Ge}(T)$). Spin relaxation time $\tau_{n-Ge}$ related to spin diffusion length by the equation $\lambda_{n-Ge} = \sqrt{D\tau_{n-Ge}}$, where $D$ is diffusion constant directly proportional to the mobility $\mu_{n-Ge}$. This allows us to plot ratio $\tau_{n-Ge}(T)/\tau_{n-Ge}(297\ K)$ as a function of temperature (Fig. 5(b)). The lowest conduction band in germanium consists of four L valleys, with center Γ valley located 0.14 eV above it. Intrinsic spin relaxation time of conduction electrons in germanium below 20 K is governed by acoustic phonons intravalley scattering between lower conduction and upper valence bands with scattering time following $T^{-7/2}$ dependence determined by Yafet [42], while above 20 K intervalley scattering between lower and upper conduction bands is dominant [43]. However, in highly doped Ge ionized impurity scattering is expected to prevail over phonon scattering, especially at low temperatures. Additionally, strain in the Ge epilayers can lift the degeneracy of the four L valleys in the lowest conduction band breaking the symmetry between them and also shifting the valence bands, thus changing respective importance of the intravalley and intervalley scattering and affecting all scattering mechanisms [43-45]. We observed saturation of the carrier mobility below 170 K, which indicates dominance of the ionized impurity scattering at low temperatures in our n-Ge channel. Spin relaxation time increased by a factor 2.2 from RT to $T = 130$ K, showing that Elliott-Yafet spin relaxation mechanism is dominant in highly doped n-Ge. From spin transport experiments Elliott-Yafet mechanism was reported to be dominant in n-Ge, but the data range was only up to 100 K [4,8], while in [9] no correlation of $\tau_{n-Ge}$ with temperature in the range from 150 K to 225 K was reported. Thus we for the first time provide clear evidences for the Elliott-Yafet spin relaxation mechanism in highly doped n-Ge from 130 K up to RT.

As a part of the control experiment, a sample of Py/n-Ge/Cu was prepared with a Cu strip (Fig. 1(b)), instead of a Me$^{SOI}$ strip. The gap distance, $L_{Py-Cu}$, was measured to be 490 nm. Figs. 6(g) and (h) show the detected electromotive force under a microwave excitation of 200 mW for two sample orientations. In contrast to the Py/n-Ge/Pt and Py/n-Ge/Pd samples, the electromotive force possesses an asymmetric shape that reverses its sign over $H_{FMR}$. This fact strongly indicates that the origin of the symmetric part of the electromotive force signal in the Py/n-Ge/Pt and Py/n-Ge/Pd samples is the ISHE in Pt and Pd, which have significantly stronger SOI than Cu. As a second part of the control experiment, the sample of Cu/n-Ge/Py/n-Ge/Pd (Fig. 1(c)) was produced with both Pd and Cu strips, which were located on different sides of the Py strip. The gap distance between the Py and Pd was measured to be $L_{Py-Pd}$ = 320 nm, while that between the Py and Cu was measured as $L_{Py-Cu}$ = 420 nm, the microwave power was set to 40 mW. From the Cu strip, similar to the previous case, an asymmetric electromotive force shape was detected (Figs. 6(i) and (j)), which is in contrast to the electromotive force from the Pd strip; the latter did not change sign over $H_{FMR}$ but instead obtained a symmetric shape (Figs. 6(k) and (l)), as expected from the ISHE. It should be noted that this result provides direct evidence of successful spin transport in n-Ge at RT because the asymmetric signal was detected from the Cu, whereas a distinctly symmetric electromotive force was detected from the Pt and Pd.

In summary, for the first time, we directly demonstrated spin transport at RT in epitaxial n-Ge with a doping concentration of ~1.0×10$^{19}$ cm$^{-3}$ using a spin pumping method and ISHE. The spin diffusion length was estimated to be $\lambda_{n-Ge} = 660 \pm 200$ nm. Spin relaxation time increased with decreasing temperature, indicating dominance of the Elliot-Yafet spin relaxation mechanism. As a result, RT spin-transport has now been shown in both pivotal semiconductor materials, Ge and Si, providing new opportunities for the future of semiconductor spintronics.

This research was supported in part by EPSRC funded "Spintronic device physics in Si/Ge Heterostructures" EP/J003263/1 and "Platform Grant" EP/J001074/1 projects.

**Figures:**

FIG. 1

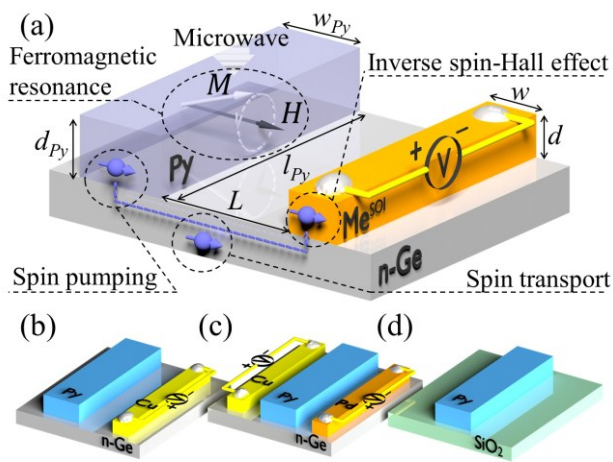

FIG. 2

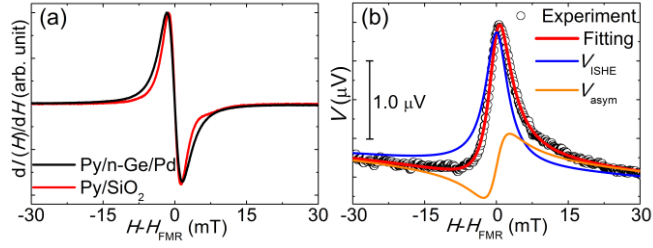

FIG. 3

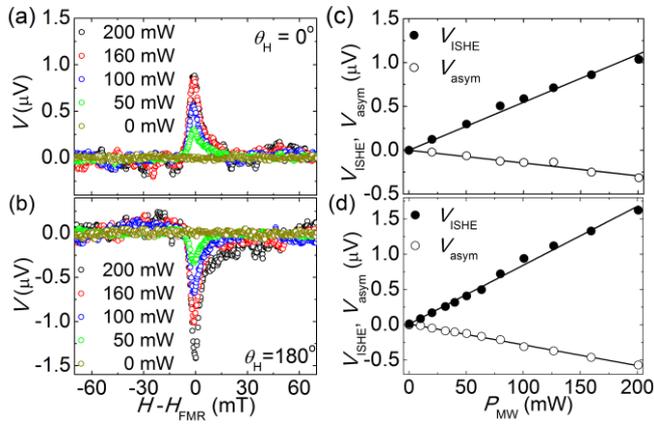

FIG. 4

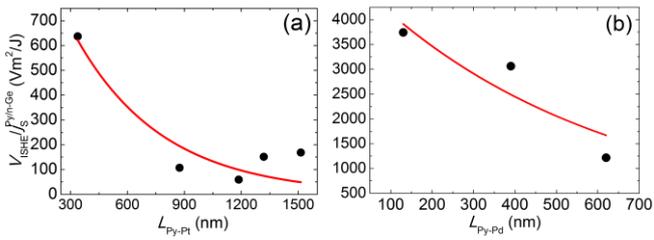

FIG. 5

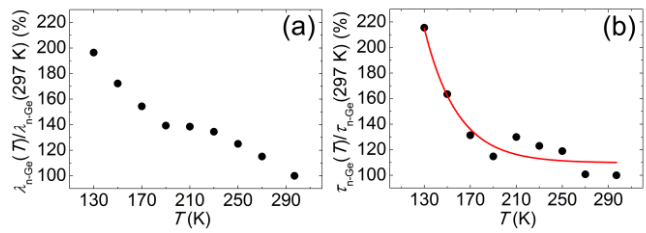

FIG. 6

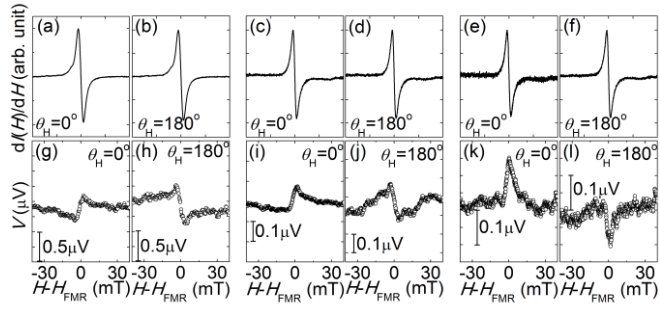

**Captions:**

FIG. 1 (color online). A schematic illustration of (a) the spin transport experiment in the Py/n-Ge/Me$^{SOI}$ device, (b) Py/n-Ge/Cu and (c) Cu/n-Ge/Py/n-Ge/Pd samples used in the control experiments, (d) Py/SiO$_2$ samples used for the spin current density estimation.

FIG. 2 (color online). (a) The FMR signal d$I$/d$H$ dependence on the in-plane external magnetic field **H** for the Py/n-Ge/Pd (black line) and Py/SiO$_2$ (red line) samples at $\theta_H = 0°$. $H_{FMR}$ and $I$ denote the resonance field and the microwave absorption intensity, respectively. (b) The electromotive force $V$ detected from the Pd strip dependence on the in-plane external magnetic field **H** for the Py/n-Ge/Pd at $\theta_H = 0°$. The open circles denote the experimental data, while the colored lines show the fitting result.

FIG. 3 (color online). The electromotive force, $V$, detected at different microwave powers from the Pt strip of the Py/n-Ge/Pt sample under the FMR for the two external magnetic field **H** orientations (a) $\theta_H = 0°$ and (b) $\theta_H = 180°$. (c,d) The microwave power dependence of the $V_{ISHE}$ (filled circles) and $V_{asym}$ (empty circles) contributions to the electromotive force $V$ detected from the Me$^{SOI}$ strip of the Py/n-Ge/Me$^{SOI}$, where the solid lines denote the linear fit. (c) Me$^{SOI}$ = Pt, (d) Me$^{SOI}$ = Pd.

FIG. 4 (color online). The gap length ($L_{Py-Me^{SOI}}$) dependence of the normalized $V_{ISHE}$ contribution to the electromotive force, $V$, detected from the (a) Pt strip of the Py/n-Ge/Pt samples and (b) Pd strip of the Py/n-Ge/Pd samples under the FMR. The filled circles represent the $V_{ISHE}$ normalized by spin current density at the Py/n-Ge interface $j_S^{Py/n-Ge}$, while the red line is the fitting curve obtained using Eq. (3).

FIG. 5 (Color online). Temperature dependence of the (a) spin diffusion length and (b) spin relaxation time of electrons in n-Ge channel normalized by RT values. Red line is exponential fitting.

FIG. 6. The FMR signal d$I$/d$H$ dependence on the in-plane external magnetic field, **H**, for the Py/n-Ge/Cu (a) $\theta_H = 0°$ and (b) $\theta_H = 180°$, for the Cu/n-Ge/Py/n-Ge/Pd (c,e) $\theta_H = 0°$ and (d,f) $\theta_H = 180°$. The electromotive force, $V$, detected under FMR from the Cu strip of the Py/n-Ge/Cu sample (g) $\theta_H = 0°$ and (h) $\theta_H = 180°$, from the Cu strip of the Cu/n-Ge/Py/n-Ge/Pd sample (i) $\theta_H = 0°$ and (j) $\theta_H = 180°$, from the Pd strip of the Cu/n-Ge/Py/n-Ge/Pd sample (k) $\theta_H = 0°$ and (l) $\theta_H = 180°$.

**An experimental demonstration of room-temperature spin transport in n-type Germanium epilayers**

S. Dushenko, M. Koike, Y. Ando, T. Shinjo, M. Myronov, and M. Shiraishi

**Calculation of the real part of the mixing conductance $g_r^{\uparrow\downarrow}$ and the spin current density at the Py/n-Ge interface $j_S^{Py/n-Ge}$.**

Under an effective magnetic field, $\mathbf{H}_{\text{eff}}$, the dynamics of the magnetization $\mathbf{M}(t)$ of the Py strip is described by Landau-Lifshitz-Gilbert equation, given as

$\frac{d\mathbf{M}(t)}{dt} = -\gamma \mathbf{M}(t) \times \mathbf{H}_{eff} + \frac{\alpha}{M_S} \mathbf{M}(t) \times \frac{d\mathbf{M}(t)}{dt}$, where, $\gamma$, $M_s$ and $\alpha$ are the gyromagnetic ratio, saturation magnetization and Gilbert damping constant, respectively. The resonance field, $H_{\text{FMR}}$, for the in-plane magnetic field is determined by $(\omega/\gamma)^2 = H_{FMR}(H_{FMR} + 4\pi M_S)$ [S1], where $\omega = 2\pi f$ is the cyclic frequency of the magnetization precession. The precession of the magnetic moment of the Py leads to spin pumping from the Py into the n-Ge with the spin current density [S2,S3] of

$$j_S^{Py/n-Ge} = \frac{\omega}{2\pi} \int_0^{\frac{2\pi}{\omega}} \frac{\hbar}{4\pi} g_r^{\uparrow\downarrow} \frac{1}{M_S^2} \left[ \mathbf{M}(t) \times \frac{d\mathbf{M}(t)}{dt} \right]_z dt; \qquad (S1)$$

here, $\hbar$ is the reduced Planck constant. The real part of the mixing conductance $g_r^{\uparrow\downarrow}$ in Eq. (S1) can be calculated from the difference in the FMR spectral width $W_{\text{Py/n-Ge}}$ of the Py/n-Ge/Pd sample relative to $W_{\text{Py}}$ of the Py/SiO$_2$ sample [S4,S5]:

$$g_r^{\uparrow\downarrow} = \frac{2\sqrt{3}\pi M_S \gamma d_{Py}}{g \mu_B \omega} (W_{Py/n-Ge} - W_{Py}), \qquad (S2)$$

where $g, \mu_B$ and $d_{Py}$ are the g-factor, Bohr magneton and thickness of the Py layer, respectively. Using the parameters $H_{FMR}$ = 96.5 mT, $g$ = 2.12 [S3], $\omega = 6.02 \times 10^{10}$ s$^{-1}$, $\gamma$ = $1.86 \times 10^{11}$ (Ts)$^{-1}$, $4\pi M_S = 0.984$ T, $d_{Py}$ = 25 nm, $W_{\text{Py/n-Ge}}$ = 3.19 mT, $W_{\text{Py}}$ = 2.55 mT, the real part of the mixing conductance was calculated to be $g_r^{\uparrow\downarrow} = 2.15 \times 10^{19}$ m$^{-2}$. Making use of the aforementioned, the spin current density at the Py/n-Ge interface is obtained to be

$$j_S^{Py/n-Ge} = \frac{g_r^{\uparrow\downarrow}\gamma^2 h^2 \hbar \left[4\pi M_S \gamma + \sqrt{(4\pi M_S)^2 \gamma^2 + 4\omega^2}\right]}{8\pi\alpha^2[(4\pi M_S)^2\gamma^2 + 4\omega^2]}; \qquad (S3)$$

here, $h$ is equal to 0.061 mT at a microwave power of 200 mW. Using Eq. (S3), $j_S^{Py/n-Ge}$ was calculated to be $1.33 \times 10^{-9}$ Jm$^{-2}$.

**Determination of the temperature dependence of the carrier mobility in the n-Ge channel $\mu_{n-Ge}$ using maximum-entropy mobility spectrum analysis.**

Hall resistance and magnetoresistance of the n-Ge channel were measured in the magnetic field range from 0 to 9 T at certain temperature. Afterwards mobility $\mu_{n-Ge}$ was determined by a maximum-entropy mobility spectrum analysis (ME-MSA) of the magnetic field dependence of the Hall resistance and magnetoresistance. Whole procedure was repeated for all needed temperatures, obtained $\mu_{n-Ge}(T)$ is shown on Fig. S1. In the ME-MSA components of the magnetoconductivity tensor $\sigma_{xy}(B)$ and $\sigma_{xx}(B)$ are calculated from the sheet values of the Hall resistivity $\rho_{xy}(B)$ and magnetoresistivity $\rho_{xx}(B)$, and then converted into mobility dependent conductivity tensor $s(\mu)$ (also called mobility spectrum) using following relations [S6,S7]:

$$\sigma_{xx}(B) = \int_{-\infty}^{\infty} \frac{s(\mu)}{1 + (\mu B)^2} d\mu, \tag{S4}$$

and

$$\sigma_{xy}(B) = \int_{-\infty}^{\infty} \frac{s(\mu)\mu B}{1 + (\mu B)^2} d\mu, \tag{S5}$$

Then the mobility spectrum $s(\mu)$ which provides the best fit to the $\sigma_{xx}(B)$ and $\sigma_{xy}(B)$ is found using entropy maximization approach. For the detailed explanation of the ME-MSA please see [S7].

**Figures:**

FIG. S1.

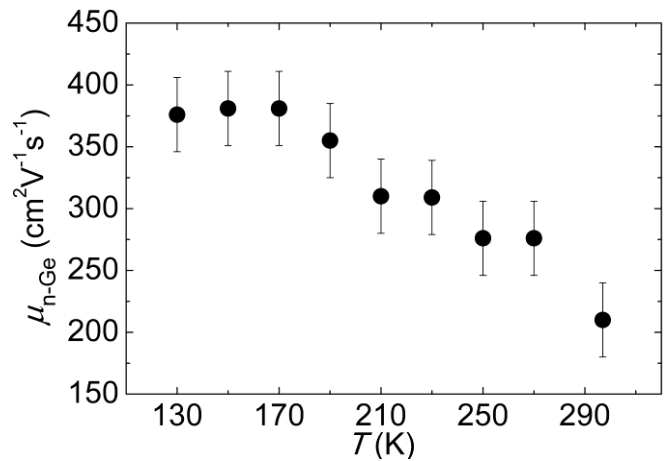

**Captions:**

FIG. S1. Temperature dependence of the mobility of the n-Ge epilayer. Error bar is defined by ME-MSA fitting parameters.